     \documentstyle[12pt]{article}   
     \newlength{\dinwidth}                       
     \newlength{\dinmargin}                      
\def\Journal#1#2#3#4{{#1} {\bf #2}, #3 (#4)}

\def\NPB{{\em Nucl. Phys.} B}
\def\PLB{{\em Phys. Lett.}  B}

\def\PRD{{\em Phys. Rev.} D}

\def\CPC{\em Computer Phys. Commun.}
\def\PR{\em Phys. Rep.}
\def\lsim{\mathrel{\rlap{\lower4pt\hbox{\hskip1pt$\sim$}}
    \raise1pt\hbox{$<$}}}                % less than or approx. symbol
\def\gsim{\mathrel{\rlap{\lower4pt\hbox{\hskip1pt$\sim$}}
    \raise1pt\hbox{$>$}}}                % greater than or approx. symbol
\newcommand{\as}{\alpha_{\mathrm{s}}}
\newenvironment{Itemize}{\begin{list}{$\bullet$}%
{\setlength{\topsep}{0.0mm}\setlength{\partopsep}{0.0mm}%
\setlength{\itemsep}{0.2mm}\setlength{\parsep}{0.2mm}}}%
{\end{list}}
\begin{document}

%set sloppy attitude to line breaks
\sloppy

\pagestyle{empty}

\begin{flushright}
LU TP 99--10 \\
hep-ph/9906316\\
June 1999
\end{flushright}
 
\vspace{\fill}

\begin{center}  \begin{Large} \begin{bf}
Some thoughts on how to match\\
Leading Log Parton Showers\\ 
with NLO Matrix Elements%
\footnote{To appear in the Proceedings of the DESY Workshop on %
Monte Carlo Generators for HERA Physics}\\
  \end{bf}  \end{Large}
  \vspace*{5mm}
  \begin{large}
C. Friberg and T. Sj\"ostrand\\
  \end{large}
 Department of Theoretical Physics, Lund University,\\ 
     Helgonav\"agen 5, S-223 62 Lund, Sweden\\ 
     christer@thep.lu.se, torbjorn@thep.lu.se
\end{center}

\vspace{\fill}

\begin{quotation}
\noindent
{\bf Abstract:}
We propose a scheme that could offer a convenient Monte Carlo
sampling of next-to-leading-order matrix elements and, at the same
time, allow the interfacing of such parton configurations with a
parton-shower approach for the estimation of higher-order effects.
No actual implementation exists so far, so this note should only be
viewed as the outline of a possible road for the future, submitted
for discussion. 
\end{quotation}

\vspace{\fill}

\clearpage
\pagestyle{plain}
\setcounter{page}{1} 

\section{Introduction}

One of the main themes of particle physics is the strive for 
an increased accuracy in the description of physical processes.
This is required to test the current standard model in detail, 
and often also to control background to searches for new physics.

A main road to improvements is the perturbative higher-order 
calculations of physical processes. For most processes, currently
this means next-to-leading order (NLO), i.e. one order higher
than the Born-level of the process, either by the emission of one 
more parton or by the inclusion of one-loop 
corrections to the Born graph. In principle, the perturbative
expansion is a well-defined and successful technique but, for 
QCD processes, the large $\as$ value makes the perturbative series 
only slowly convergent. This problem is especially severe in the 
collinear region, where the emission rate is increasing 
as one approaches the non-perturbative r\'egime. Finite total cross
sections are obtained only by a cancellation between large positive
real and large negative virtual contributions. Therefore higher-order 
matrix elements (ME's) are not of much use to describe the 
substructure of jets, apart from very crude features. 

Parton showers (PS's) have complementary strengths. By a resummation
of the large logarithmic terms, e.g. into Sudakov form factors,
it is possible to obtain a reasonable 
description also in regions of large $\as$ values. Formally, for 
most generators, this resummation is only certified to leading
logarithmic (LL) accuracy, but in reality many of the expected
next-to-leading log improvements are already included, such as
exact energy-momentum conservation, angular ordering, and optimal 
scale choice for $\as$. Furthermore, the cross section for any 
$n$-parton configuration is always positive definite.
Finally, it is possible to terminate the parton showers at some
process-independent lower cut-off $Q_0$ and attach a --- thereby 
also process-independent --- non-perturbative hadronization model 
for physics below that scale. 

The main PS weakness, on the other hand, is the crude treatment of 
wide-angle parton emission, where many Feynman diagrams may contribute 
with comparable strengths, and the final rate therefore may depend on 
detailed interference effects not present in the PS language. For
some simple processes it has been possible to improve the showers by 
explicit NLO ME information in this region, thereby obtaining an 
improved description of the process \cite{existingmatches}. In
general, however, this approach does not appear tractable, and
one would like to find other methods to combine the advantages
of the ME answer at large parton separations with the PS one at
small separation (and with hadronization models for scales below
that).

In this note we will present some thoughts on a more general
--- although maybe less beautiful --- ME/PS matching strategy that 
could be used for a larger set of NLO processes. To be specific, we
will consider the example where the leading-order process is of the
$2 \to 2$ type, i.e. producing two high-$p_{\perp}$ jets, as observed
in HERA photoproduction events. The NLO corrections then contain both   
$2 \to 3$ processes and virtual corrections to the $2 \to 2$ ones.

\section{The NLO parton configurations}

The first step is to use the ME's to set up the starting configuration,
either 2- or 3-``jet'' events, where ``jet'' is a misnomer denoting that
any nearby partons have been clustered. The two event classes are 
distinguished by some jet resolution criteria, i.e. by requiring that 
none of the partons in a $2 \to 3$ configuration are found in a soft 
or collinear region. This can be defined, e.g., in terms of minimal 
$p_{\perp}$ and $R = \sqrt{\Delta\eta^2 + \Delta\phi^2}$, or minimal 
invariant masses between partons. 

While it is straightforward to `cut out' the appropriate phase 
space regions from the 3-``jet'' final state, to remain with a finite 
and positive differential cross section everywhere else, the consequences 
for the 2-``jet'' configurations are more complicated. Here one will 
now receive contributions from\\
(1) the leading-order $2 \to 2$ ME's,\\
(2) the virtual terms, including (counter)terms coming from the
scale-dependent parton distributions, also $2 \to 2$, and\\
(3) those $2 \to 3$ parton configurations that are rejected as
such by the criteria above, and thus should be reclassified as 
2-``jet'' events.

Divergences from soft and collinear emissions should cancel
between the two latter event classes. In the extreme divergent
region, the three-body phase space naturally reduces to the two-body 
one, and so the singularities can be cancelled analytically. The
finite pieces are often integrated numerically as three separate
contributions, that are only combined in the end to give the 
correct cross section. If this strategy is applied in an event 
generator, it becomes necessary to work with events with negative
weight, from the virtual corrections term. This is possible, but 
known to be a very unstable procedure in practice. For instance, 
it requires the hadronization model to be continuous in the limit 
that two partons are brought closer and eventually merged to one.
This is true for the string model \cite{AGIS}, although some fine
print sets a practical limit, but many other hadronization models 
are flawed in this respect from the onset.

Instead we would propose a rearranged procedure, inspired by the 
so-called subtraction method \cite{LEP2}. (Which does not exclude 
the use of the competing phase-space slicing method to handle the 
collinear regions.) In analogy with this method, the third term is 
subdivided in two:\\
(3a) a strongly simplified matrix element, that reproduces the 
correct behaviour in the singular regions, but away from this
can be chosen in a convenient way that allows simple integration
over the extra 3-body phase space variables not present in the
2-body phase space (for fixed incoming partons), and\\
(3b) the difference between the full and the approximate 
expressions, that is everywhere finite, but messy to integrate
analytically.
 
In a Monte Carlo context, this would work as follows:
\begin{Itemize}
\item Pick a desired $2 \to 2$ parton configuration, at random
(but biased to the regions of large cross sections, of course).
\item Evaluate the differential cross section contributions from 
the parts (1) and (2+3a) for this configuration. By the notation 
(2+3a) we imply that the singular contributions now explicitly 
cancel between (2) and (3a). We expect (1)+(2)+(3) to be clearly
positive, in the sense that, were (2)+(3) anywhere to become negative 
of the same order as (1) is positive, one would be entering a 
collinear/soft region of large higher-order corrections, better 
described by showers. Therefore the parameters of the clustering 
algorithm must be chosen so as to avoid this. Provided that the 
approximate form in (3a) is not very badly chosen, also (1)+(2+3a) 
should always be positive, although this is not strictly required.
\item Pick a 3-body phase-space point in the soft/collinear
regions that, by the jet resolution criteria, should be classified
as the 2-``jet'' configuration picked above. 
\item Evaluate the difference (3b) in this point, and multiply    
by the integral of the extra 3-body phase space variables.
\item Add (1)+(2+3a)+(3b) together to obtain the cross section 
for the two-``jet'' configuration. With a reasonable separation 
into (3a) and (3b), the grand sum should always be a non-negative 
number, that can then be used to accept/reject events in order to 
obtain a final event sample with unit weight. 
\end{Itemize} 
The key feature here is that, by the Monte Carlo nature of it,
one and the same two-``jet'' configuration will be assigned 
different (non-negative) weights each time the cross section is 
evaluated, since the associated three-parton configuration will differ, 
but it is arranged so that the average converges to the right cross 
section.   

\section{The PS interface}

A parton shower is organized in terms of some evolution variable,
such that emissions are ordered to give e.g. a decreasing angle,
transverse momentum or mass \cite{LEP2}. Often the upper limit of 
evolution is set to cover the full phase space of emissions, but some 
other maximum can also be indicated. For instance, if the ME 
regularization has been defined in terms of some minimum invariant 
mass (or angular) scale between resolved partons, a natural complement 
is a shower evolution in mass (or angle) from this scale downwards. 
However, such a match would never be perfect, so one would always need 
the capability to reject unwanted branchings in a shower. Fortunately,
an acceptance/rejection step is part of shower algorithms anyway,
so one only needs to add further rejection criteria matching the 
NLO ME cuts.

The method would therefore be as follows:
\begin{Itemize}
\item In two-``jet'' events, the evolution is started from some 
conveniently large scale, chosen so that no allowed phase-space
regions are excluded. When a potential new emission has been
selected by the shower algorithm, the resulting new parton 
configuration is tested by the ``jet'' clustering algorithm. 
Any emission that gives a three-``jet'' classification is rejected 
and the evolution is continued downwards. This scheme should be 
applied both for the initial- and final-state showers. Ambiguities
could arise, e.g. if one emission from the initial and one from the 
final state happen to overlap, so that they together define a third
``jet'', although individually they do not. Occurrences of this kind 
are formally of higher order, so one is free to pick any sensible
strategy. One extreme would be only to apply cuts to each shower 
separately, another to insert a final clustering test that makes use 
of all partons to accept/reject the full shower treatment.    
\item In three-``jet'' events, in principle all further emission 
is allowed, also such that leads to four or more ``jets''. However,
if one were to allow emissions at scales harder than the ones in
the basic $2 \to 3$ graph itself, there is a manifest risk of 
doublecounting in the jet cross section. Again, this could be avoided
by applying a veto to emissions. In \cite{Andre} an explicit
algorithm is presented, that sets up a final-state shower from a given 
parton configuration, in a sensible way that avoids (or at least
minimizes) doublecounting. A similar approach should be possible for 
the initial-state showers.
\end{Itemize}  

\section{Outlook}

If we want to improve the precision of NLO QCD tests in the future,
new strategies need to be developed. This note is one such proposal.
The main point is an alternative phase-space sampling/integration 
strategy for NLO ME's. It would lead to events with positive definite 
weights, that therefore could be better interfaced to parton showers 
and hadronization models, and thus more realistically compared
with data. 

Clearly many details need to be settled, to make this a working
proposition. While it should not be necessary to recalculate any
of the NLO corrections to a process, the code for the evaluation
of  cross sections need to be restructured compared with
current practice. Especially, the phase-space generation machinery
must be rewritten significantly. By comparison, the modifications
required in existing parton-shower algorithms appear more straightforward.

\noindent
{\bf Acknowledgment:}We thank M. Klasen and B. P\"otter for helpful 
conversations.

\end{document}